\documentclass[preprint,showpacs,preprintnumbers,amsmath,amssymb]{revtex4}
\usepackage{graphicx,textcomp}
\usepackage{dcolumn}
\usepackage{bm}
\usepackage{amsmath}

\begin{document}

\title{Potential energy and dipole moment surfaces of H$_3^-$ molecule.}
\author{M. Ayouz$^{1}$, O. Dulieu$^{1}$, R. Gu\'erout$^{1}$, J. Robert$^{1}$, and V. Kokoouline$^{1,2}$}
\affiliation{$^{1}$Laboratoire Aim\'e Cotton, CNRS, B\^at. 505, Universit\'e Paris-Sud, 91405 Orsay Cedex, France\\
$^{2}$Department of Physics, University of Central Florida, Orlando, Florida 32816, USA }

\begin{abstract}
A new potential energy surface for the electronic ground state of the simplest triatomic anion H$_3^-$ is determined for a large number of geometries. Its accuracy is improved at short and large distances compared to previous studies. The permanent dipole moment surface of the state is also computed for the first time. Nine vibrational levels of H$_3^-$ and fourteen levels of D$_3^-$ are obtained, bound by at most $\sim 70$~cm$^{-1}$ and  $\sim 126$~cm$^{-1}$ respectively. These results should guide the spectroscopic search of the H$_3^-$  ion in cold gases (below 100K) of molecular hydrogen in the presence of H$^-$ ions.
\end{abstract}

\pacs{82.20.Kh, 34.20.-b, 82.30.Fi }

\maketitle

\section{Introduction}

Collisions involving hydrogen atoms, molecules, and their positive (H$^+$, H$_2^+$) and negative (H$^-$) ions play an important role in chemistry and evolution of neutral or negatively-charged hydrogen plasma such as laboratory hydrogen plasma, the interstellar medium (ISM),  atmospheres of the Sun and other stars \cite{rau96} as well in the Earth atmosphere.  Binary collisions of these species are simple enough to be treated using first principle methods without any adjustable parameter. Therefore, such processes are often used as benchmarks for testing theoretical methods.

The existence of bound states of the H$_3^-$ ion in a linear configuration for the three nuclei has been suggested in 1937 \cite{stevenson37}, but has been questioned since then.  Different {\it ab initio} calculations  \cite{vanderslice60,ritchie68,macias68a,macias68b,garcia79,chalasinski87,michels87,starck93,panda04} have been giving contradictory results for the depth of the van der Waals well. The estimated error bar in the calculations was comparable or larger than obtained binding energies, so the calculations  could not predict for sure if the H$_3^-$ ion is stable.   Although first observations of H$_3^-$ ions have been reported in low resolution experiments as early as in 1974 \cite{hurley74,aberth75,schnitzer78,bae84,wang03,golser05}, they could not warrant either the stability of H$_3^-$. It is only in the 1990's \cite{starck93,robicheaux99} that theory became precise enough to confirm the stability of H$_3^-$ bound states. 

At present, there are three available accurate potential energy surfaces (PES) of H$_3^-$: The PES by St\"arck and Meyer \cite{starck93}, by Belyaev and Tiukanov \cite{belyaev97}, and by  Panda and Sathyamurthya \cite{panda04}. In the following, these three PES will be referred to as PES-SM, PES-BT, and PES-PS respectively. In addition, Belyaev, Tiukanov and others \cite{kabbaj88,belyaev93,belyaev97,belyaev99,belyaev01,belyaev09}  have obtained PES of excited electronic states of H$_3^-$ and their non-Born-Oppenheimer couplings with the ground state. The excited electronic states are unstable with respect to electron autodetachment. Bound state calculation based on the PES-SM have been performed for H$_3^-$, H$_2$D$^-$, and D$_2$H$^-$ in Ref. \cite{starck93} and using the PES-BT for H$_3^-$ in Ref. \cite{wang03}.  It is worth to mention a stand alone study by Robicheaux \cite{robicheaux99} confirming the stability of H$_3^-$ (and similar anions), where the PES was derived from the scattering length for the electron-H$_2$ collisions, and from the polarizability of H$_2$.

While never observed in the ISM, the collisions between H$_2$ and H$^-$, and of their isopotologues, have been studied in a number of laboratory experiments back to the 1950's \cite{muschlitz56,muschlitz57,huq83,zimmer92,zimmer95,muller96,haufler97}. Such collisions  should also play a role in processes in tokamaks, especially if the negative ion source  (D$^-$) is used for the tokamak neutral beam injectors \cite{okumura00,grisham01}.  Many studies have been devoted to the calculation of elastic and inelastic cross-sections for H$_2$-H$^-$ collisions and for all isotopologues,  \cite{macon58,mahapatra95,mahapatra96,gianturco95,ansari98,belyaev00,mahapatra00,jaquet01,panda04, panda05,giri06,yao06,giri07,belyaev09}, most of them for energies significantly larger than 1~eV above the lowest dissociation limit H$_2$ + H$^-$. However, at such energies the theoretical cross-sections may not be reliable:  the dissociation energy of the H$_3$ ground state lies only 0.75~eV above the H$_3^-$ dissociation energy, and the non-Born-Oppenheimer interaction between the ground states of H$_3$ and H$_3^-$ could be significant.  The non-Born-Oppenheimer interactions have been taken into account only in Ref.  \cite{aguillon00} within a reduced-dimensionality approach, where only linear geometries of H$_3^-$ have been taken into account. Finally, the photodissociation of H$_2$D$^-$ has been studied in a simplified approach based on the Franck-Condon overlap between a bound  level and a scattering state of  H$_2$D$^-$ \cite{takayanagi00}, since the {\it ab initio} dipole moments of H$_3^-$ were not available at that time.

The present study is mainly motivated by the formation of H$_3^-$ bound states in low-energy collisions between H$_2$ and H$^-$.  In such collisions H$_3^-$ can be formed only if a third body (other than  H$_2$ or H$^-$) participates (three-body recombination -- TBR) or if a photon is emitted (radiative association -- RA). Both processes could be relevant for the chemistry of cold interstellar clouds, if H$^-$ is present \cite{kokoouline10}. Note that the H$^-$ ion has not been detected so far in the interstellar medium (ISM): it cannot be directly observed by usual photoabsorption spectroscopy because  H$^-$ has only one bound electronic state. 

The evaluation of the cross-section for radiative association of H$_2$ and H$^-$ at low energy (10-30K) requires an accurate PES with a precision around 1 cm$^{-1}$ or better, and permanent dipole moment surfaces (PDMS) for H$_3^-$. In the present study we calculated a new  {\it ab initio} PES for the H$_3^-$ electronic ground state on a dense and large grid for internal coordinates, using a larger electronic basis set than those of PES-SM and PES-PS calculations. Special care is taken in order to account for the long-range behavior of the surface. We also obtain for the first time the PDMS of the H$_3^-$ electronic ground state. We constructed Fortran subroutines that calculate PES and PDMS values for any arbitrary geometry using B-spline interpolation procedures, and we determine the bound states of H$_3^-$ and D$_3^-$. The RA reaction will be treated in a separate study.

The article is organized in the following way. In the next section, we discuss the calculation of the new PES and the interpolation procedure. In Section \ref{sec:comparison} we compare the new PES with the PES from previous studies. Section \ref{sec:vib_levels} is devoted to the calculation of bound levels of H$_3^-$ and  D$_3^-$ and Section \ref{sec:dipole} presents our results on the dipole moment of H$_3^-$. Section \ref{sec:concl} is the conclusion. Atomic units (a.u.) for distances (1 a.u. = 0.0529177~nm) and for energies (1~a.u. = 219474.63137~cm$^{-1}$) will be used throughout the paper, except otherwise stated.

\section{{\it Ab initio} calculation and interpolation of the H$_3^-$ ground state potential surface}
\label{sec:abinitio}

As in Ref. \cite{starck93}, we used the coupled-electron pair approximation (CEPA-2) method \cite{meyer73}, which is part of the Molpro package \cite{MOLPRO_brief}. It is a non-variational variant of the configuration interaction method  for closed-shell molecules. Here we used a considerably larger basis set,  AV5Z with $spdfg$ basis functions from the Molpro basis library, and a much larger number of geometries, than in Refs \cite{starck93,panda04}. As H$_3^-$ is a van der Waals molecule, we defined a 3D-grid in Jacobi coordinates: $r$-- the distance between two protons, $R$-- the distance from the center of mass of the two protons to the third proton, and $\gamma$ -- the angle between vectors $\vec R$ and $\vec r$. The grid in $r$ is uniform from $r=0.8$ a.u. to $r=2.4$ a.u. and changes by a step $\Delta r=0.2$ a.u. The grid in $\gamma$ changes from $0$ to 90\textdegree~  by a constant step of $\Delta \gamma=$15\textdegree. The grid in $R$ was chosen denser for small $R$ than for large $R$: the grid points $R_i$ were calculated according to  $R_i=1.5+0.452\exp(i/10)$ (in a.u.) with $i=1,2,\cdots, 48$, which makes $R$ changing from 1.9995 a.u. to 56.4227 a.u. Therefore, the calculations were performed for $9\times 7\times 48=3024$ geometries. Notice that the above grid starts at a smaller value of $r=0.8$ a.u. than in Refs. \cite{starck93,panda04} ($r=1$ a.u.). Indeed, we found that a grid starting at $r=1$ cannot properly represent the repulsive part of the H$_2$ ground state potential curve (at fixed $R$ and $\gamma$) and gives an appreciable  error in  energies obtained of the lowest H$_2$ vibrational levels and, as a result, a comparable error in dissociation energies of H$_3^- \to$H$_2(v,j)+$H$^-$ when $R\to\infty$. We also used a denser grid in $\gamma$ and a longer and denser grid in $R$ than in  Refs. \cite{starck93,panda04}.

\begin{figure}[ht]
\includegraphics[width=12cm]{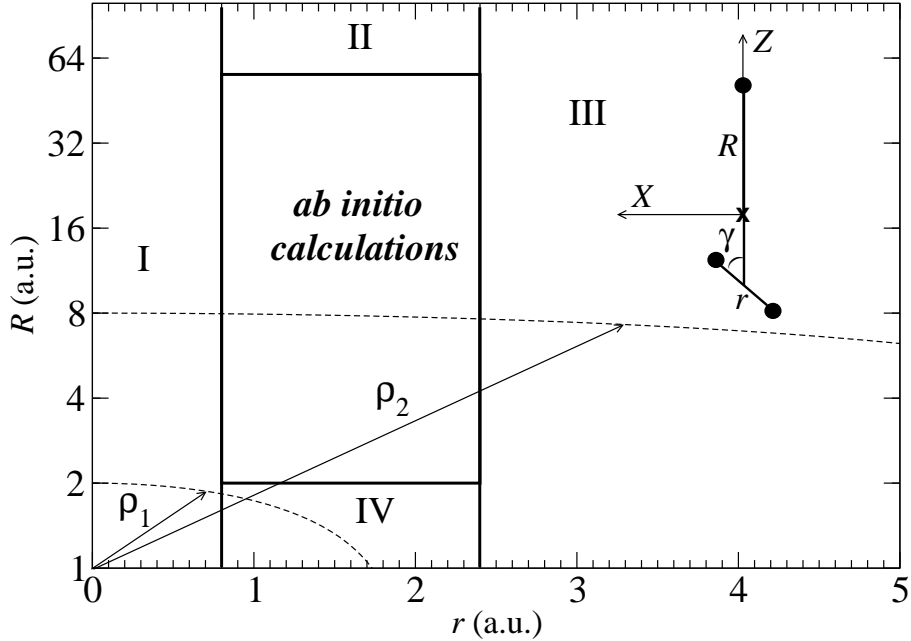}
\caption{The five regions of the configuration space in Jacobi coordinates $R,r$ defined for the PES calculations, for all values of $\gamma$ (see the text).  Notice that $R$ is given on a logarithmic scale. The central rectangle represents the region of calculated \textit{ab initio} points, where a 3D B-spline interpolation is used.  Different methods were used for the extrapolation of the PES and PDMS in regions I-IV (see text).  The arcs of a circle (dashed lines) represent lines of fixed hyperradii $\rho$ (see Section \ref{sec:vib_levels}) for two values, $\rho_{1}=2$ a.u. and $\rho_{2}=8$ a.u. The $X$,  $Y$, and $Z$ axes  (needed to specify the components of the PDM vector)  are the axes of principal moments of inertia. Their orientation is  indicated in the figure. The $Y$-axis is out the plane of H$_3^-$.}
\label{fig:Grid_Extrapolation}
\end{figure}

For each geometry we calculated the potential energy of the H$_3^-$ ground state, and the components of the PDM vector with respect to the principal axes of inertia. The obtained \textit{ab initio} PES and PDMS were used to prepare Fortran subroutines that calculate PES and PDMS for any arbitrary geometry. Inside the three-dimensional box  $r\in [0.8;2.4],\ R\in [1.9995;56.4227]$, $\gamma\in[0;360$\textdegree] (central region in Fig. \ref{fig:Grid_Extrapolation}) the procedures interpolate the surfaces using the three-dimensional B-spline method. Outside of the box, the procedures rely on analytical formulas for the extrapolation of PES and PDMS.
To simplify the description of this region, we divided the $r\times R$ configuration space in four different parts (I-IV) surrounding the {\it ab initio} rectangle (Fig. \ref{fig:Grid_Extrapolation}). Regions ``{\it ab initio calculations}`` and II are  the only ones relevant to bound and scattering states of the system with energies $\lesssim 2$eV above the  H$_2(v=0,j=0)$+H$^-$ dissociation. Therefore, the extrapolation procedure for the PES and PDMS in the two regions should be physically justified. In contrast, wave functions of bound and continuum states with such energies vanish in regions I,  III, and IV, and empirical formulas will be used. As it will be discussed in Section \ref{sec:vib_levels}, the extrapolation is needed to map the configuration space in Jacobi coordinates on the space of hyperspherical coordinates, which will be used for bound state and scattering calculations.

In region II, we represent the long-range (in $R$, at fixed $r$ and $\gamma$) potential $V_{LR}$ for the interaction between  H$_{2}$ and H$^-$  as \cite{buckingham67}:
\begin{eqnarray}
\label{eq:VLR_R}
V_{LR}(R;r,\gamma)= D_{as}(r)+\frac{C_3^\textrm{th}}{R^3}+\frac{C_4^\textrm{th}}{R^4}\ \textrm{with}\\
C_3^\textrm{th}=-Q(r)P_{2}(\cos\gamma)\ \textrm{and}\ C_4^\textrm{th}=-\left[\alpha_{0}(r)+\alpha_{2}(r)P_{2}(\cos\gamma)\right]/2\nonumber\,,
\end{eqnarray}
where the first term $D_{as}(r)$ is the sum of H$^-$ and H$_{2}(r)$ energies at a given internuclear distance $r$ of the H$_2$ molecule. The second term is the interaction between the electric charge of H$^-$ with the quadrupole moment $Q(r)$ of H$_2$ (taken from Ref. \cite{poll78}), and the third term is the interaction of the dipole moment of H$_2$ induced by H$^-$, involving the second order Legendre polynomial $P_{2}(\cos \gamma)$.  The functions $\alpha_{0}(r)$ and $\alpha_{2}(r)$ are the isotropic and anisotropic polarizabilities of H$_{2}$, for which we used the analytical functions given in Ref. \cite{panda04} that were obtained by fitting the numerical values from Ref. \cite{kolos67}. The dispersion energy varying as $1/R^6$ and other smaller terms are neglected, which is a good approximation because the long-range expansion is only used for $R>56.4$ a.u.

In region I we used the following extrapolation formula in $r$ for fixed $R$ and $\gamma$:
\begin{eqnarray}
\label{eq:VSR_r}
V_{SR}(r;R,\gamma)=a(R,\gamma)e^{-b(R,\gamma)r}\,,
\end{eqnarray}
where  $a(R,\gamma)$  and $b(R,\gamma)$ are functions of $R$  and $\gamma$ that are obtained considering the two {\it ab initio} energies $V(r=0.8;R,\gamma)$ and $V(r=1;R,\gamma)$ calculated at first two values of the coordinate $r$. In this way, we obtain the quantities $a$ and $b$ given on a two-dimensional grid of points  in the ($R$,$\gamma$) space. Then we used the 2D B-spline interpolation to obtain smooth two-dimensional functions $a(R,\gamma)$ and $b(R,\gamma)$. 


In region III, we extrapolate the PES in $r$ and at fixed $R$ and $\gamma$ using a dispersion-like expression: 
\begin{eqnarray}
\label{eq:VLR_r}
V_{LR}(r;R,\gamma)=D_{0}(R,\gamma)-\frac{C_{6}(R,\gamma)}{r^6}\,,
\end{eqnarray}
where the $D_{0}(R,\gamma)$ and $C_{6}(R,\gamma)$ (always positive) coefficients are obtained in a way similar to the coefficients $a$ and $b$, considering the two last points $V(r=2.2$ a.u.$ ;R,\gamma)$ and $V(r=2.4$ a.u.$;R,\gamma)$. They are also interpolated using the 2D B-spline method for arbitrary values of $R$ and $\gamma$.

The behavior of the PES in region IV is described by the short-range (in $R$) repulsive expression at given values of $r$ and $\gamma$:
\begin{eqnarray}
\label{eq:VSR_R}
V_{SR}(R;r,\gamma)=A(r,\gamma)e^{-B(r,\gamma)R}\,,
\end{eqnarray}
The $A(r,\gamma)$ and $B(r,\gamma)$ coefficients are obtained in a way similar to the one described above, considering the first two values of the {\it ab initio} potential energy $V(R=2$ a.u.$;r,\gamma)$ and $V(R=2.052$ a.u. $;r,\gamma)$, and are further interpolated for any ($r$, $\gamma$) using the same 2D B-spline procedure.

\begin{figure}[ht]
\includegraphics[width=7cm]{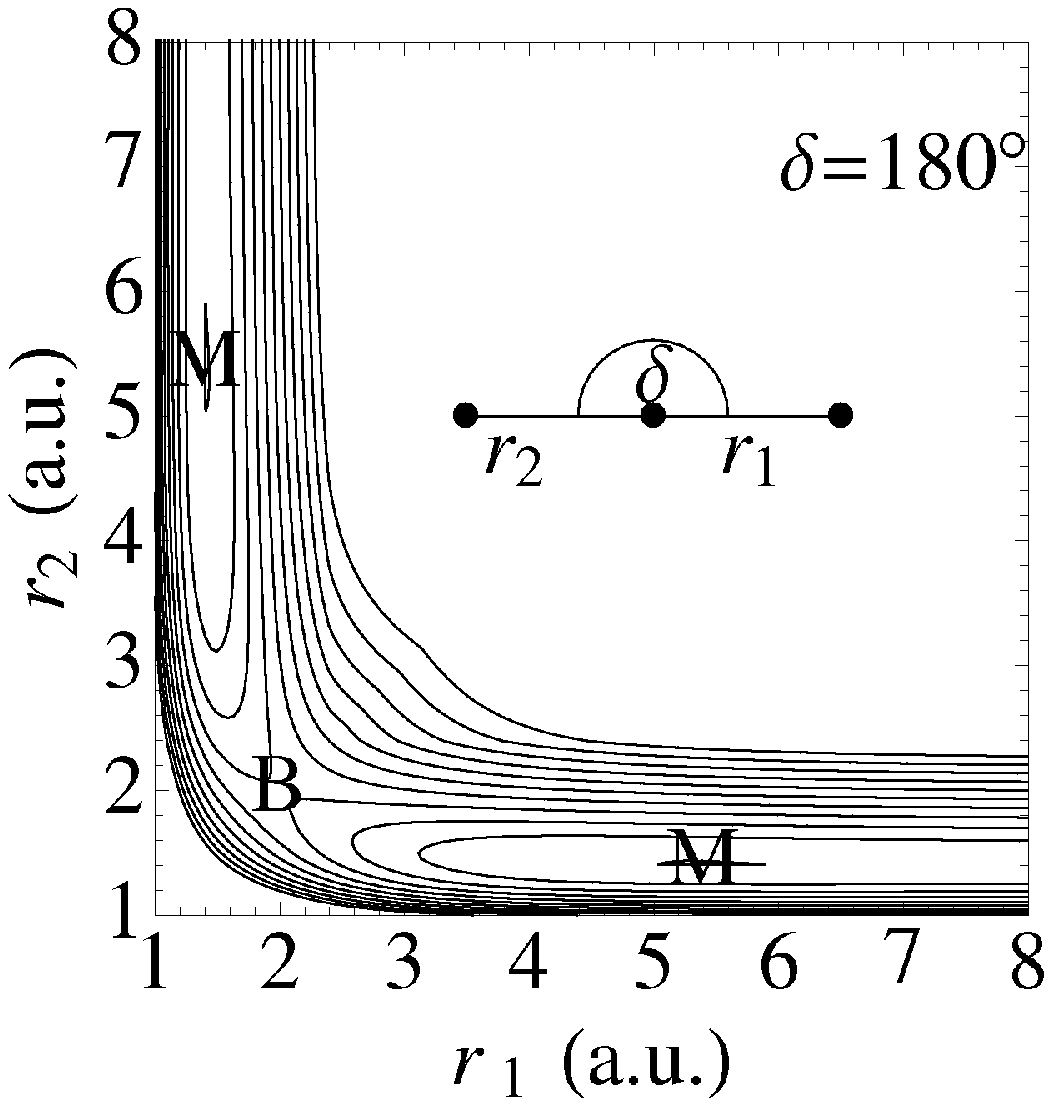}\qquad
\includegraphics[width=7cm]{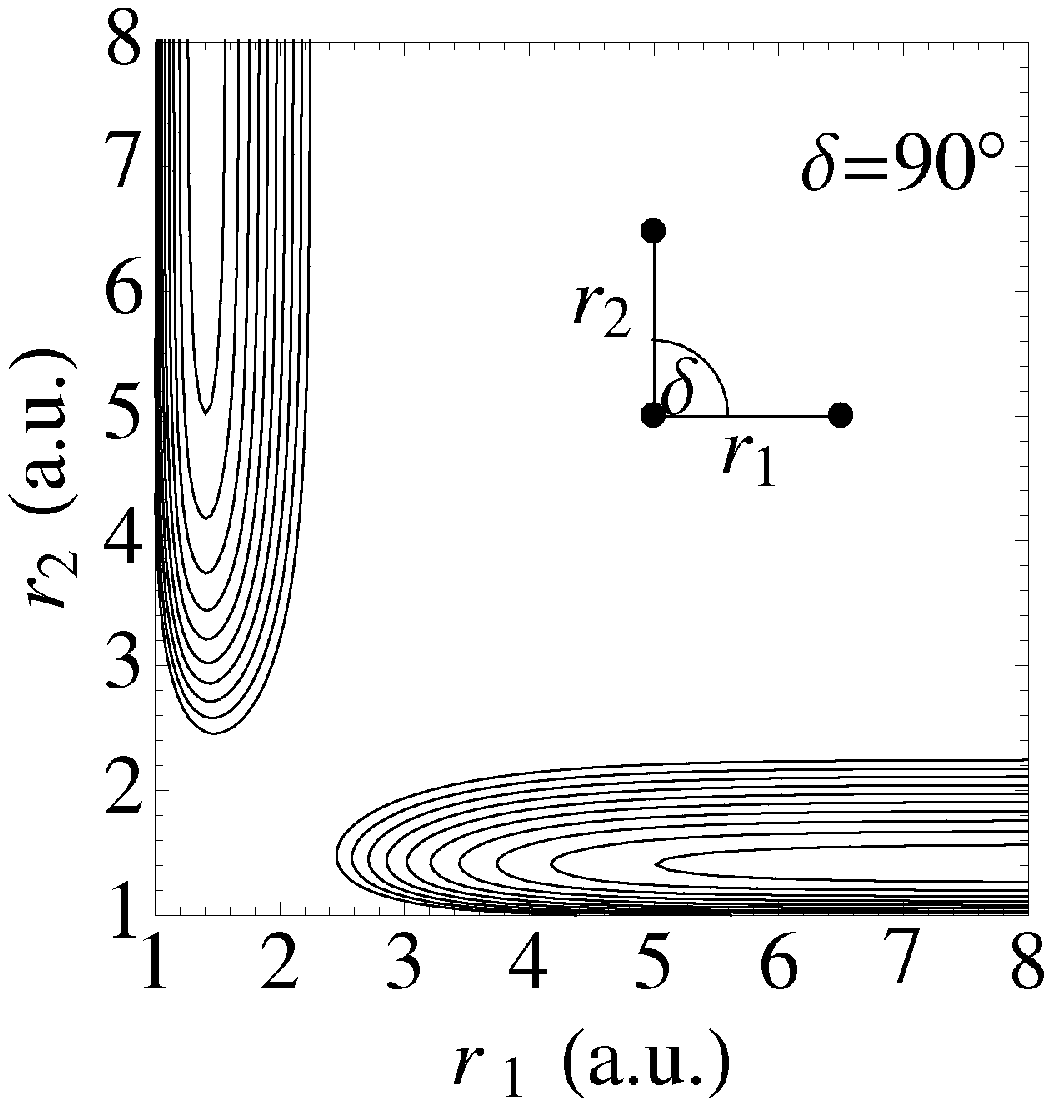}
\caption{Energy contour plots of the H$_3^-$ electronic ground state PES in the space of the two internuclear distances ($r_{1}$ and  $r_{2} $) for two values of the bonding angle  $\delta$ defined in each plot. Successive contours differ by $0.006$ a.u. The dissociation energy $D_{as}=-1.7016828$ a.u. of the PES corresponds to the infinite separation between H$_{2}(r_{e}$=1.403 a.u.) and H$^-$ (see Table \ref{tab:SummaryTab}). The energy minimum (indicated by $M$) is obtained for a linear configuration, and the top of the potential barrier is labeled with $B$. The lowest contour corresponds to the energy of $-1.7034$ a.u. (left plot) and $-1.6974$ a.u. (right plot).}
\label{fig:MolProPES}
\end{figure}

We show in Fig. \ref{fig:MolProPES} the H$_3^-$ PES as a function of two internuclear distances, $r_{1}$ and  $r_{2} $ for  two values (180 and 90\textdegree) of the bonding angle $\delta$. These coordinates are convenient to describe main features of the PES. As expected, the figure is symmetric with respect to the  $r_{1}\leftrightarrow r_{2}$ exchange, and shows the van der Waals  potential well in each coordinate (with the minimum labeled with $M$), separated by a potential barrier (with the maximum labeled with $B$) for the exchange of two identical nuclei. As in previous studies, lowest energy is found for a linear configuration. The position and energy of $M$ and $B$ are reported in  Table \ref{tab:SummaryTab}, together with other characteristic parameters of the PES.

The Fortran subroutines calculating the PES and PDMS from the present {\it ab initio} data using the interpolation and extrapolation procedures described above are available with the journal's EPAPS service and can be provided by the authors upon a request.

\begin{table}[tbp]
\begin{tabular}{|c|c|c|c|c|}
\hline
Quantity & Present study & Ref. \cite{starck93} & Ref. \cite{panda04}&Ref. \cite{belyaev97}\\
\hline
$D_{as}^{a)}$  (a.u.)& -1.701683  & -1.70095 & n/a &n/a\\ 
\hline
$r_{e}^{b)}$(a.u.)& 1.403  & 1.40  & n/a &n/a\\ 
\hline
$B^{c)}$ (a.u.)& -1.68509& -1.68562 & n/a&n/a$^{d)}$\\ 
\hline
position of $B$  & &  & &\\ 
$r_{1}$, $r_{2} $ (a.u.),  $\delta=0$ & 1.996, 1.996  &  1.997 , 1.997& 1.999, 1.999& 1.74, 1.74\\ 
$r,R$(a.u.),  $\gamma=0$ & 1.996, 2.994  &  1.997, 2.996& 1.999, 2.999& 1.74, 2.61\\ 
\hline
$M^{e)}$ (a.u.) & -1.703511 & -1.70270 & n/a&n/a\\
\hline
position of $M$  & &  & &\\ 
$r$, $R$ (a.u.),  $\gamma=0$     & 1.421, 6.069& 1.416, 6.183 & 1.419, 5.915 &n/a\\ 
\hline
$B-M$ (cm$^{-1}$)& 4042.9   & 3748.62   & 3786.42  &n/a\\
\hline
$D_{as}-M$ (cm$^{-1}$) & 401.2 & 384.05 & 384.27  & 443.60\\
\hline
\hline
\multicolumn{5}{|c|}{Energies obtained in calculation made separately for H$_{2}$ and H$^-$}\\
\hline
$E_{\text{H}_{2}(r=1.4)}$&-1.174252& -1.17368& n/a&n/a\\
\hline
$E_{\text{H}^-}$&-0.527429& -0.52727& n/a&n/a\\
\hline
\end{tabular}
\caption{Different quantities characterizing the H$_3^-$ ground state PES obtained in the present study compared with previous calculations \cite{starck93,panda04,belyaev97}. ${a)}$ asymptotic energy at infinite separation between H$_{2}(r_{e})$ and H$^-$; ${b)}$ internuclear distance corresponding to the minimum of the H$_2$ dimer potential; ${c)}$ energy of the maximum of the barrier, $B$ in Fig. \ref{fig:MolProPES}; ${d)}$ Ref.  \cite{belyaev97} gives the height of the barrier (0.624 eV) with respect to the H$_{2}(v=0,j=0)$+H$^-$ dissociation; ${e)}$ energy of the PES minimum, $M$ in Fig. \ref{fig:MolProPES}.}
\label{tab:SummaryTab}
\end{table}

\section{Comparison with previous studies}
\label{sec:comparison}

\begin{figure}[ht]
\includegraphics[width=10cm,angle=0]{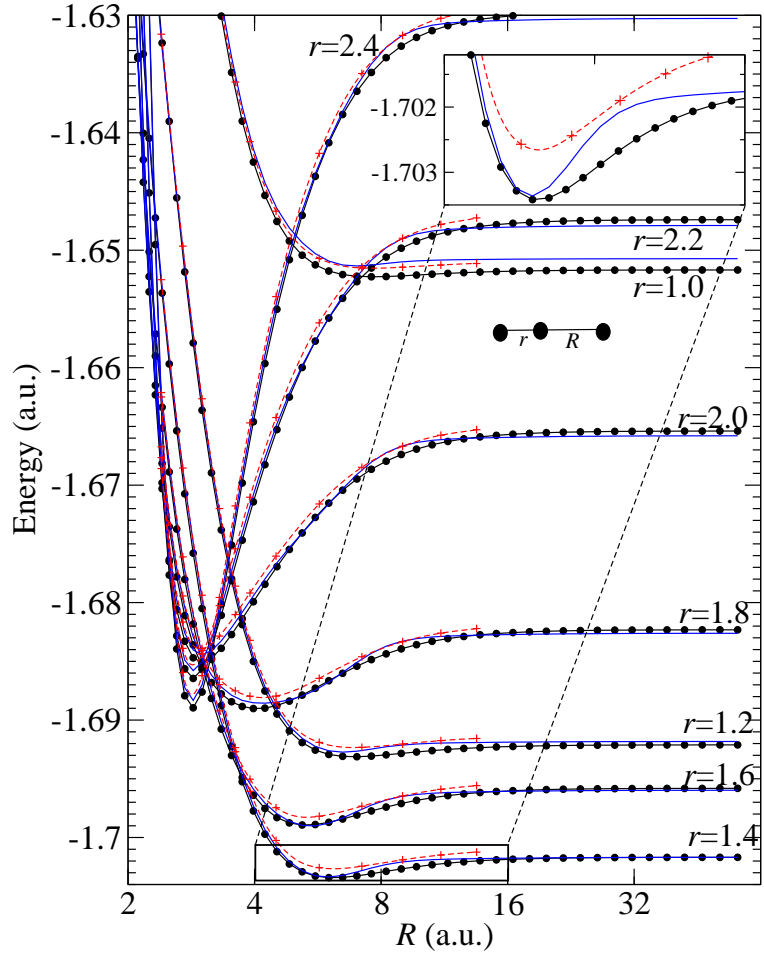}
\caption{(Color online) Comparison of the present H$_3^-$ PES (circles) in Jacobi coordinates with the results from Ref. \cite{starck93} (crosses on the dashed lines) and Ref.  \cite{panda04} (solid blue lines with no symbols) for $\gamma=0$\textdegree. The PES are shown as a function of $R$ for eight different values of $r$. The $r=0.8$ a.u. curve is not shown here because it is located at a higher energy.  Symbols indicate the geometries for which the actual {\it ab initio} calculations have been performed. The continuous lines connecting the points are obtained by interpolation. No  {\it ab initio} energies are available for the PES of Ref. \cite{panda04}: the solid lines without symbols are obtained directly from the analytic formula and the Fortran subroutine provided in Ref. \cite{panda04}. Notice the logarithmic scale in $R$ variable. The inset enlarges the region of the minimum of the PES close to equilibrium region at $r=1.4$ a.u.}
\label{fig:Meyer_Panda_MolPro_0}
\end{figure}

\begin{figure}[ht]
\includegraphics[width=12cm,angle=0]{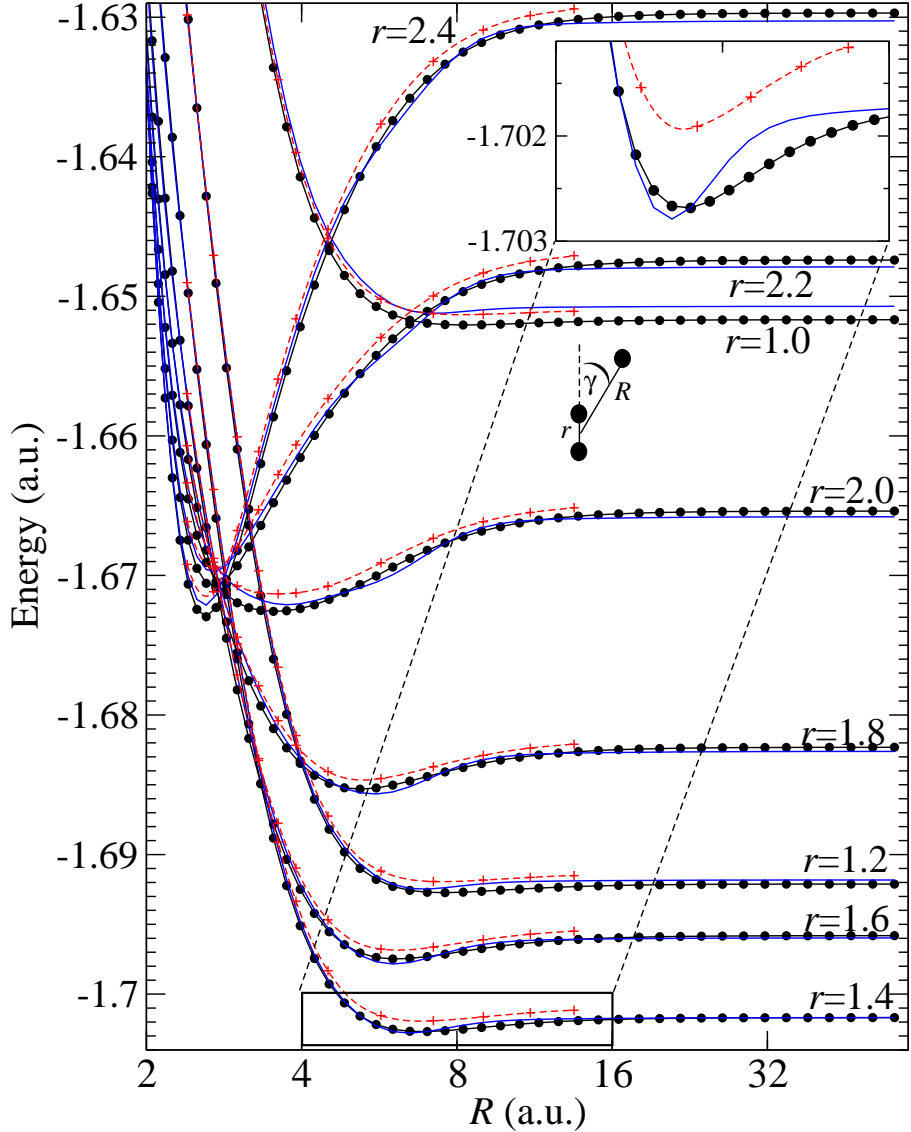}
\caption{(Color online) Same as Fig. \ref{fig:Meyer_Panda_MolPro_0} except $\gamma=30$\textdegree. }
\label{fig:Meyer_Panda_MolPro_30}
\end{figure}

\begin{figure}[ht]
\includegraphics[width=12cm,angle=0]{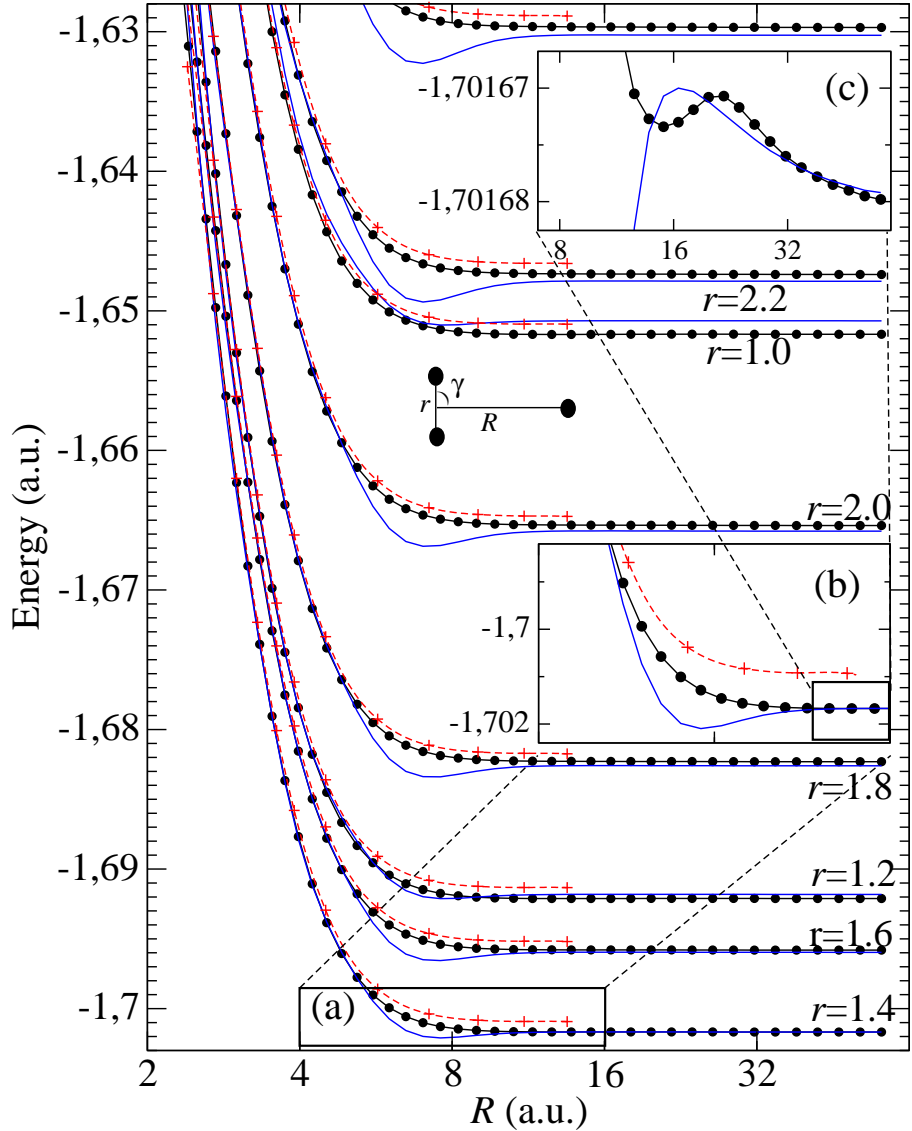}
\caption{(Color online) Same as Fig. \ref{fig:Meyer_Panda_MolPro_0} except $\gamma=90$\textdegree. The two insets, (b) and (c), show the behavior of the potentials for $r=1.4$ a.u. on two different magnification scales:  Region (b) is the magnified version of (a), region (c) is  magnified  (b).}
\label{fig:Meyer_Panda_MolPro_90}
\end{figure}

Figures \ref{fig:Meyer_Panda_MolPro_0}, \ref{fig:Meyer_Panda_MolPro_30}, and 
 \ref{fig:Meyer_Panda_MolPro_90} illustrate the comparison of the PES in Jacobi coordinates obtained in the present study with the results of Refs. \cite{starck93,panda04} for three values of $\gamma=$0\textdegree, 30\textdegree, and 90\textdegree\ respectively.  Each figure gives the PES for one value of $\gamma$ and eight values of $r$ versus the coordinate $R$. 
 The origin of  potential energy yielded by the {\it ab initio} procedure corresponds to an infinite separation of all electrons and nuclei. Because the absolute energy of the  PES of Ref. \cite{panda04} is unknown, the origin of this surface is chosen in such a way that the asymptotic energy of the infinite separation between H$_{2}(r=$1.4 a.u.) and H$^-$ is the same as in the present study. 

The insets of these figures, together with the data in Table \ref{tab:SummaryTab} demonstrate that the present {\it ab initio} energies are systematically lower that those of  Ref. \cite{starck93} for all geometries. For instance, the potential well depth has  increased by 0.01\% (around 160~cm$^{-1}$) at $r_e=1.4$ a.u. This was expected, as we use the same CEPA-2 method than in Ref. \cite{starck93} with a larger basis set, leading then to a PES with an improved accuracy. The overall behavior of the present PES and PES-SM as a function of  $R$ is similar at small and large distances. In contrast, a noticeable difference of PES-PS with the present results and the PES-SM is the presence of a potential well at $\gamma=90$\textdegree\  near  $R=7.6$ a.u. (see the inset in Fig. \ref{fig:Meyer_Panda_MolPro_90}) with a significant depth ($\sim$90 cm$^{-1}$ for $r=1.4$ a.u.). For  $\gamma=90$\textdegree\ the $C_3/R^3$ long-range energy contribution is positive and $C_4/R^4$ is negative (see Eq. \ref{eq:VLR_R}). Therefore, their sum combined with the short-range interaction produces a potential curve (for fixed $r$ and  $\gamma$ near 90\textdegree\ ) that has a pronounced minimum in PES-PS and just a should-like feature in the present PES (see inset (c) in Fig. \ref{fig:Meyer_Panda_MolPro_90}). The curve  is repulsive asymptotically. A possible reason for the above differences is the somewhat smaller basis set (d-aug-cc-PVTZ) used in calculation of PES-PS \cite{panda04}.

The numerical data available about PES-BT \cite{belyaev97,belyaev01}  is given in Table \ref{tab:SummaryTab}. Since the calculation method (diatomics-in-molecule --DIM) explicitly uses the {\it ab initio} data (wave functions and energies) obtained for fragments (H$_2$, H$^-$, H$_2^-$, H$_3$) in separate  {\it ab initio} calculations, it  gives automatically the correct dissociation energy.

Since we have calculated PES at relatively large distances $R$, we can extract {\it ab initio} values (labelled with ''ai``) of the corresponding long-range coefficients $C_3^\mathrm{ai}$ and  $C_4^\mathrm{ai}$ and compare them with the theoretical values $C_3^\mathrm{th}$ and $C_4^\mathrm{th}$ of Eq. (\ref{eq:VLR_R}). We first make a fit of $C_4^\mathrm{ai}$ to the {\it ab initio} data by fixing $C_3^\mathrm{ai}$ to the theoretical value $C_3^\mathrm{th}$. In this way the value of $C_4^\mathrm{ai}$ agrees with the value of $C_4^\mathrm{th}$ within 0.3\% for $\gamma=0$, and 5 \% for $\gamma=90$\textdegree at $r=1.4$ a.u. Then, we did the opposite:  we fitted  $C_3^\mathrm{ai}$ to the {\it ab initio} data by fixing $C_4^\mathrm{ai}$ to $C_4^\mathrm{th}$. In this case we found that fitted and theoretical  values of $C_3$ agree within 0.1 \% for $\gamma=0$ and 2 \% for $\gamma=90$\textdegree at $r=1.4$ a.u. The PES-PS differs significantly from the analytical long-range behavior of Eq. (\ref{eq:VLR_R}). Surprisingly, the long-range behavior of the PES-PS is quite different from ours and the one of PES-SM, and probably inaccurate due to the matching procedure between the long and short distances used by these authors.

\section{Vibrational states of H$_{3}^-$ and D$_{3}^-$}
\label{sec:vib_levels}

As mentioned in the introduction, one of the motivations of this study is to describe low-energy collisions between H$_2$ and H$^-$, and to investigate the formation of H$_3^-$ in such collisions. Therefore we need an approach which treat both bound and continuum states of the triatomic anion (including rearrangement of nuclei). We use the Smith-Whitten hyperspherical coordinates \cite{johnson80}: 
a hyper-radius $\rho$ and two hyperangles $\theta,\ \phi$, which can be defined for the three identical particles by the three internuclear distances $r_i,\ (i=1,2,3)$  as
\begin{eqnarray}
r_i= 3^{-1/4}\rho\sqrt{1+\sin\theta\sin(\phi+\epsilon_i)}\,,
\end{eqnarray}
where $\epsilon_1=2\pi/3$, $\epsilon_2=-2\pi/3$, and $\epsilon_3=0$. We also employ the adiabatic separation between hyperradius and hyperangles, which are known to be well adapted  to atom-molecule inelastic and reactive scattering involving identical particles. The dynamics is  treated within the framework of the slow variable discretization (SVD) method \cite{tolstikhin96,kokoouline06,blandon07} that allows us to account easily for non-adiabatic couplings between hyperspherical adiabatic channels. We briefly recall below the main steps of our approach, which is discussed in greater details in Refs. \cite{kokoouline06,blandon07}. 

The eigenstates $\Psi$ of three particles interacting through a potential $V$ depending only on the three internuclear distances are obtained by solving the Schr\"odinger equation with the following Hamiltonian  expressed in hyperspherical coordinates $\rho, {\omega}$ in the center-of-mass frame \cite{johnson80}:
\begin{equation}
-\frac{1}{2\mu}\rho^{-5}\frac{\partial}{\partial\rho}\rho^5\frac{\partial}{\partial\rho}+\frac{\Lambda^2}{2\mu\rho^2}+V\,,
\end{equation}
where $\mu=m/\sqrt{3}$ is the three-body reduced mass and $m$ is the mass of each of the three identical particles. The operator $\Lambda$ above is the  grand angular momentum \cite{smith60b,johnson83b}. It depends only on the set ${\omega}$ of five angles, which include the three Euler angles (for the orientation of the molecular frame in the lab frame) and the hyperangles $\theta$ and $\phi$. If the total angular momentum $J=0$, $\Lambda$ depends only on the two hyperangles. The explicit form of $\Lambda^2$ is given, for example, by Eq. (27) of Ref.  \cite{johnson83b}:
\begin{eqnarray}
\Lambda^2=-\frac{4}{\sin(2\theta)}\frac{\partial}{\partial\theta}\sin(2\theta)\frac{\partial}{\partial\theta}-\frac{4}{\sin^2(\theta)}\frac{\partial^2}{\partial\phi^2}\\
+\frac{2J_X^2}{1-\sin\theta}+\frac{2J_Z^2}{1+\sin\theta}+\frac{J_Y^2}{\sin^2\theta}+\frac{4i\cos\theta J_Y }{\sin^2\theta}\frac{\partial}{\partial\phi }\,,  \nonumber
\end{eqnarray}
where $J_X$, $J_Y$, and $J_Z$ are  the components of the angular momentum along the principal axes of inertia.  The orientation of the axes is approximately indicated in Fig. \ref{fig:Grid_Extrapolation}. After rescaling the wave function $\Psi$ as $\Psi=\rho^{-5/2}\psi$, the Hamiltonian for the new function $\psi$ is written as
\begin{eqnarray}
H=T_\rho+H_\textrm{ad}\,,\ \textrm{where}\\
T_\rho=-\frac{1}{2\mu}\frac{\partial^2}{\partial\rho^2}\ \textrm{and}\\
 H_\textrm{ad}=\frac{\Lambda^2+15/4}{2\mu\rho^2}+V\,.
\end{eqnarray}

The eigenfunctions $\psi(\rho,{\omega})$ of the above Hamiltonian are sought as an expansion over the basis functions $y_{a,j}(\rho,{\omega})$ with unknown coefficients $c_{a,j}$:
\begin{equation}
\label{eq:expansion}
 \psi(\rho,{\omega})=\sum_{a,j}  y_{a,j}(\rho,{\omega})c_{a,j}\,
\end{equation}
The basis functions $y_{a,j}(\rho,{\omega})$ are constructed as products
\begin{equation}
y_{a,j}(\rho,{\omega})= \pi_j(\rho)\varphi_{a,j}({\omega})\,
\end{equation}
where the  functions $\pi_j(\rho)$ are DVR-like functions localized at DVR grid points $\rho_j$ along the hyper-radius. As in our earlier study  \cite{kokoouline99,kokoouline00b}, here we used the plane wave DVR functions. The functions $\varphi_{a,j}({\omega})$ are the adiabatic hyperspherical states obtained by solving the three-body Schr\"odinger equation at fixed $\rho=\rho_j$, {\it i.e.} they are eigenstates of $H_\textrm{ad}$ at fixed $\rho=\rho_j$ with eigenvalues $U_a(\rho_j)$
\begin{eqnarray}
\label{eq:ad-states}
H_\textrm{ad}^{\rho=\rho_j}\varphi_{a,j}({\omega})=U_a(\rho_j)\varphi_{a,j}({\omega})\,.
\end{eqnarray}

The functions $U_a(\rho_j)$ are usually referred to as hyperspherical adiabatic potentials.
Inserting the expansion of Eq. (\ref{eq:expansion}) into the Schr\"odinger equation $H\psi=E\psi$ reduces the equation to a generalized eigenvalue problem for coefficients $c_{a,j}$ with eigenvalues $E$
 \begin{eqnarray}
\label{eq:gen_eigen_value}
  \sum_{a'j'}\left[\left\langle\pi_{j'}\arrowvert -\frac{1}{2\mu}\frac{\partial^2}{\partial\rho^2}\arrowvert \pi_j\right\rangle_\rho{O}_{a'j',aj}+\langle\pi_{j'}\arrowvert U_a(\rho)\arrowvert\pi_j \rangle_\rho\delta_{a'a} \right]c_{j'a'}=\nonumber\\
E \sum_{a'j'}\langle\pi_{j'}\arrowvert\pi_j \rangle_\rho{O}_{a'j',aj}c_{j'a'} \,,
 \end{eqnarray}
where ${O}_{a'j',aj}$ are the overlap integrals (in the ${\omega}$ space) between adiabatic states $\varphi_{a',j'}$ and $\varphi_{a,j}$:
\begin{eqnarray}
{O}_{a'j',aj}=\langle\varphi_{a',j'}\arrowvert\varphi_{a,j}\rangle_{\omega}\,.
 \end{eqnarray}
The subscripts $\rho$ and ${\omega}$ at kets in the above expressions refer to the integration coordinate of the bracket. 

The system  (\ref{eq:gen_eigen_value}) of equations resembles to the system of coupled-channel equations, where non-adiabatic couplings $\langle\varphi_{a'}\lvert\frac{\partial}{\partial\rho}\rvert\varphi_{a}\rangle\frac{d}{d\rho}$ and $\langle\varphi_{a'}\lvert\frac{\partial^2}{\partial\rho^2}\rvert\varphi_{a}\rangle$ are replaced with the overlap matrix elements ${O}_{a'j',aj}$. The use of overlap matrix elements instead of the derivatives of adiabatic states with respect to $\rho$ simplifies significantly the numerical solution of the equation \cite{kokoouline06,blandon07}, and is the main advantage of the SVD method.

We restricted the present computations to energies and wave functions of bound states for two H$_3^-$ and D$_3^-$, with total angular momentum $J=0$. The hyperangle $\theta$ varies in the interval [$0$,$\frac{\pi}{2}$] while the full interval of variation for the second hyperangle, $\phi$, is [0,2$\pi$). However, for three identical particles, the interval along $\phi$ can be restricted to [$-\frac{\pi}{2}$,$-\frac{\pi}{6}$] for wave functions of the $A_1'$ or $A_2'$ irreducible representations (\textit{irreps} in the following) of the molecular symmetry group ($D_{3h}$), and to the interval [$-\frac{\pi}{2}$,$+\frac{\pi}{2}$] for wave functions of the $E'$ \textit{irrep}. \textit{Irreps} with odd parities, $A_1''$, $A_2''$, and $E''$ are not allowed for $J=0$. This is because the inversion applied to the rotational part of the total wave function is reduced to the rotation of the system about the $Y$ axis by $\pi$ \cite{bunker98}. The $J=0$ rotational states are isotropic, so only the even parity is allowed. The hyperradius $\rho$ can vary in the interval [$0$,$\infty$). In this study, it varies from 1 a.u. to 120 a.u. for H$_3^-$ a.u. and from 1 a.u. to 80 a.u. for D$_3^-$. In the numerical calculation, we have used equal masses $m=1837.3621$ a.u. for all three atoms in the H$_{3}^-$ molecule. This value is the sum of the hydrogen mass and one third of electron mass. Similarly, we took $m=3670.8162$ a.u. (which is the sum of the deuterium mass and one third of electron mass) for D$_3^-$. 

\begin{figure}[ht]
\includegraphics[width=12cm]{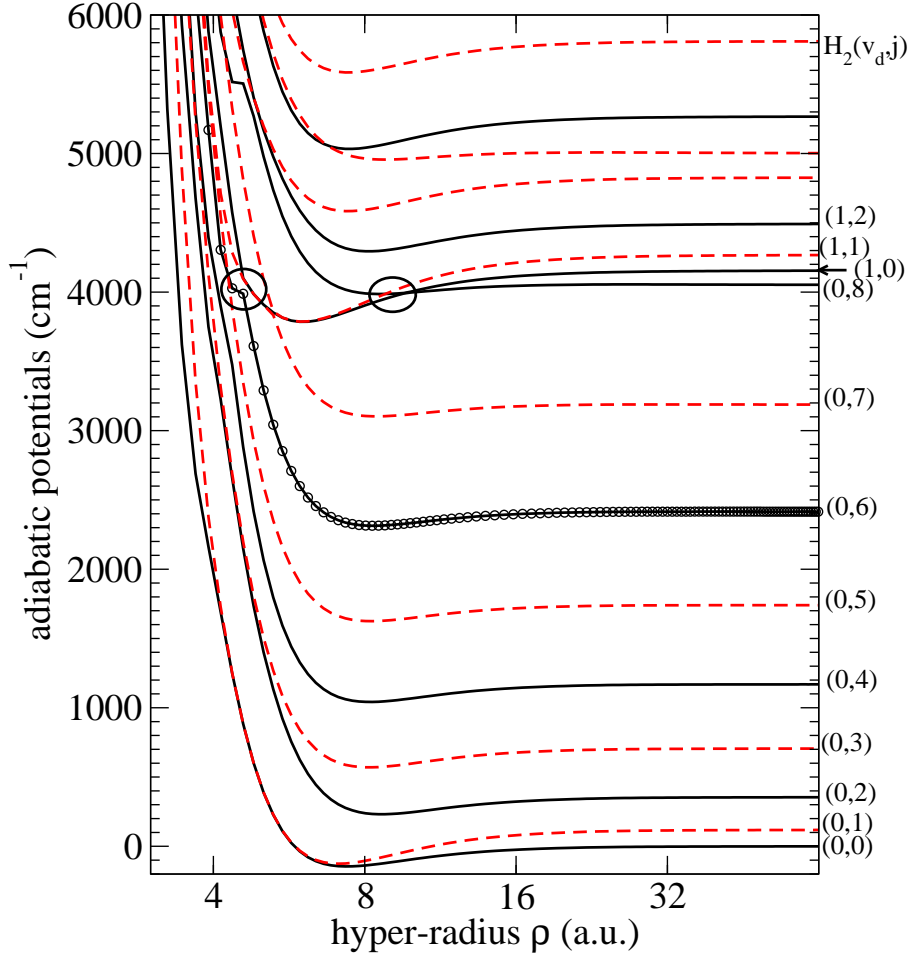}
\caption{(Color online) The lowest H$_{3}^-$ hyperspherical adiabatic curves ($U_a(\rho)$ in Eq. (\ref{eq:ad-states})) for $J=0$ of $A_1'$ \textit{irrep} (solid black lines) and of $A_2'$ \textit{irrep} (dashed red lines). The curves of the $E'$ \textit{irrep} are not displayed, as they are very similar to the $A_{1}'$ and $A_{2}'$ curves for energies smaller than 4000~cm$^{-1}$ above the lowest dissociation limit H$_{2}(0,0)+\text{H}^-$, and would be indistinguishable on this scale. A couple of avoided crossings located around 4000~cm$^{-1}$ are marked. Each curve is labelled at large $R$ with the pair of indexes ($v_d, j$) corresponding to the rovibrational state of the H$_2$ molecule. The energy origin is set to the lowest dissociation limit H$_{2}(0,0)+\text{H}^-$. The closed circles on the (0,6) curve exemplifies the density of grid points in $R$ used in the calculations.}
\label{fig:H3_Adiabatic_Curves}
\end{figure}

\begin{figure}[ht]
\includegraphics[width=12cm]{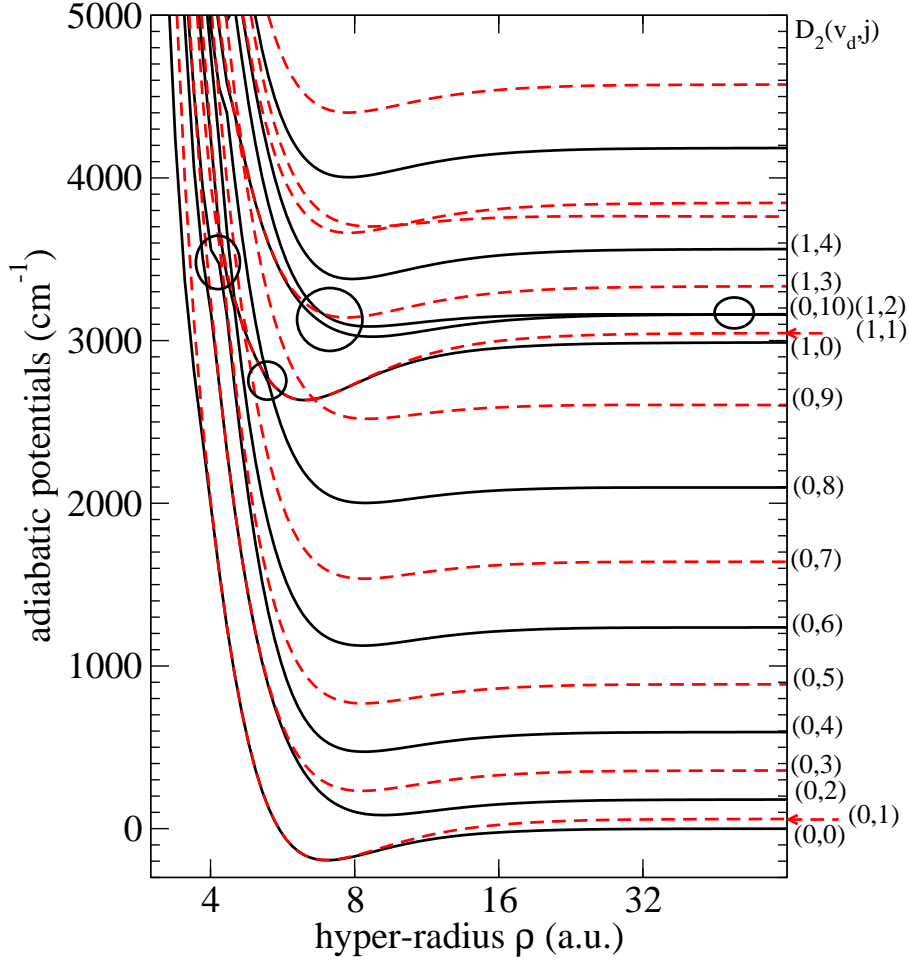}
\caption{(Color online) Same as Figure \ref{fig:H3_Adiabatic_Curves} for the D$_3^-$ molecule. The energy origin here is set to the lowest dissociation limit D$_{2}(0,0)+\text{D}^-$. A few avoided crossings are indicated by circles.}
\label{fig:D3_Adiabatic_Curves}
\end{figure}

The resulting adiabatic potentials of $A_1'$ and $A_2'$ \textit{irreps} are shown in Fig. \ref{fig:H3_Adiabatic_Curves} for H$_{3}^-$ and in Fig. \ref{fig:D3_Adiabatic_Curves} for D$_{3}^-$. The curves of $E'$ \textit{irrep} are not shown in Figs. \ref{fig:H3_Adiabatic_Curves} and \ref{fig:D3_Adiabatic_Curves}: The lowest $E'$ curves are almost identical to the $A_1'$ and $A_2'$ curves and would be indistinguishable from them in the figures because for low vibrational levels, the energies of levels are independent on the symmetry ($A_1',A_2'$, or $E'$)  of wave functions  with respect to the proton exchange between the dimer and the H$^-$ ion.  (We assume here that the dimer is in a particular rovibrational state such that the proton exchange is made without changing the dimer state. The energy does depend on the H$_2$ rovibrational state).  At large $\rho$, all adiabatic curves dissociate into an atom+dimer system characterized by the rovibrational state $(v_\mathrm{d},j)$ of the dimer. 

Some of the adiabatic curves in Figs. \ref{fig:H3_Adiabatic_Curves} and \ref{fig:D3_Adiabatic_Curves} exhibit avoided crossings at energies above $\sim 4000$ cm$^{-1}$: Dynamic coupling between hyperspherical adiabatic states is mostly determined by the avoided crossings and responsible for nuclei exchange above the potential barrier identified in Fig. \ref{fig:MolProPES}. Such transitions are much less probable at smaller energies as each adiabatic state is only weakly coupled to other adiabatic states of the same \textit{irrep}. As a consequence, only one component $\varphi_a$ in the expansion of Eq. (\ref{eq:expansion}) is dominant. 
Each adiabatic state for a given \textit{irrep} $\Gamma$ is correlated with a definite pair of quantum numbers $(v_\mathrm{d},j)$, and can be approximately characterized by these two quantum numbers. The value $\Omega$ of the projection of the H$_2$ angular momentum $\mathbf{j}$ on the axis connecting the dimer with the atom is also a relevant quantum number. For $J=0$, $\Omega$ is always zero and, therefore is not specified in Figs. \ref{fig:H3_Adiabatic_Curves} and \ref{fig:D3_Adiabatic_Curves}. 

To characterize completely a bound state of the trimer, an additional quantum number $v_t$ is needed for the excitation within each adiabatic state (along the hyperradius). Therefore, the bound states are characterized by four approximate quantum numbers $\Omega,j,v_t,v_d$ and two exact quantum numbers $J$ and $\Gamma$. At high energies, the mixing between different $\Omega,j,v_t,v_d$ for given $J$ and $\Gamma$ becomes important.

We summarized in Table \ref{tab:H3_BoundStates} the present H$_{3}^-$ bound state energies and those of Ref. \cite{starck93} labelled with the set of quantum numbers $J$,$j$,$\Omega$,$v_{t}$,$v_{d}$,$\Gamma$. In agreement with Ref. \cite{starck93}, we found four vibrational energies levels of the $A_{1}'$ \textit{irrep} and one other level of the $A_{2}'$ \textit{irrep}. We also found four more vibrational levels of the $A_{2}'$ \textit{irrep}. Although their energies are located above the $v_d=0,j=0$ dissociation threshold, they are stable because they cannot dissociate due to the symmetry restriction. Their binding energies are given in the table with respect to their first allowed dissociation limit  $v_d=0,j=1$. It is worth noticing that our binding energies for all levels but one are larger than the energies of Ref. \cite{starck93}), expressing that the well depth of our potential surface is found about 17~cm$^{-1}$  deeper than in Ref. \cite{starck93}). The last bound level of the $A_{1}'$ \textit{irrep} has a very small binding energy and may not be fully converged. The bound state energies of D$_{3}^-$ are given in Table \ref{tab:D3_BoundStates}. Since the D$_3^-$ molecule is heavier than H$_3^-$, it has more vibrational levels. No data from previous calculation is available for D$_{3}^-$.

\begin{table}[htp]
\begin{tabular}{|p{2.5cm}|p{3cm}|p{3cm}|p{3cm}|p{3cm}|}
\hline
& \multicolumn{2}{|c}{Energies above $D_{as}$} & \multicolumn{2}{|c|}{Binding energies}\\
\hline
\multicolumn{1}{|c|}{J,$\Omega$,j,$v_{\text{t}}$,$v_{\text{d}}$,$\Gamma$}& \multicolumn{1}{c|}
{Present study} &\multicolumn{1}{c|}{Calculations of \cite{starck93}}& \multicolumn{1}{c|}{Present study}
 &\multicolumn{1}{c|}{Calculations of \cite{starck93}}\\
\hline
0,0,0,0,0,$A_{1}'$ & 2103.3& 2100.6 &-70.7 &-68.4\\
0,0,0,1,0,$A_{1}'$ & 2148.4& 2147.1 &-25.6 &-21.9\\
0,0,0,2,0,$A_{1}'$ & 2168.7& 2165.9 &-5.4  &-3.1\\
0,0,0,3,0,$A_{1}'$ & 2174.1& 2168.7 &-0.01 &-0.3\\
$D_{00}$\footnote{$^)$ Asymptotic energy of  H$^-$+H$_{2}(v_{d}=0,j=0)$ dissociation}$^)$& 2174.1& 2169.0&&\\
0,0,1,0,0,$A_{2}'$ & 2140.3& 2140.8 &-152.1 (-33.7) &n/a (-28.2)\\
0,0,1,1,0,$A_{2}'$ & 2215.1& &-77.4  &\\
0,0,1,2,0,$A_{2}'$ & 2259.3& &-33.2  &\\
0,0,1,3,0,$A_{2}'$ & 2281.9& &-10.6  &\\
0,0,1,4,0,$A_{2}'$ & 2290.9& &-1.6   &\\
$D_{01}$\footnote{$^)$ Asymptotic energy of  H$^-$+H$_{2}(v_{d}=0,j=1)$ dissociation}$^)$& 2292.5 & n/a&&\\
\hline
\end{tabular}
\caption{Comparison of energies (in cm$^{-1}$) of H$_3^-$ bound levels obtained in the present study with those of Ref.  \cite{starck93}. Energies in the second and third columns are given with respect to $D_{as}$ (see Table \ref{tab:SummaryTab}). The binding energies (fourth and fifth columns) are given with respect to the dissociation limits $D_{00}$ (for $A_1'$ levels) and $D_{01}$ (for $A_2'$ levels). The energies in parentheses are given with respect to  $D_{00}$, to compare with Ref.\cite{starck93}.}
\label{tab:H3_BoundStates}
\end{table}

\begin{table}[htp]
\begin{tabular}{|p{2.5cm}|p{4cm}|p{4cm}|}
\hline
\multicolumn{1}{|c|}{J,$\Omega$,j,$v_{\text{t}}$,$v_{\text{d}}$,$\Gamma$}& \multicolumn{1}{c|}{Energies above $D_{as}$}& 
\multicolumn{1}{c|}{Binding energies}\\
\hline
0,0,0,0,0,$A_{1}'$ & 1416.5& -126.2 \\
0,0,0,1,0,$A_{1}'$ & 1474.1& -68.6 \\
0,0,0,2,0,$A_{1}'$ & 1510.1& -32.6 \\
0,0,0,3,0,$A_{1}'$ & 1530.2& -12.5 \\
0,0,0,4,0,$A_{1}'$ & 1539.8& -2.9 \\
0,0,0,5,0,$A_{1}'$ & 1542.7& -0.02 \\
$D_{00}$\footnote{$^)$ Asymptotic energy of  D$^-$+D$_{2}(v_{d}=0,j=0)$ dissociation}$^)$& 1542.7& \\
0,0,1,0,0,$A_{2}'$ & 1419.9&-182.3 \\
0,0,1,1,0,$A_{2}'$ & 1485.0&-117.2 \\
0,0,1,2,0,$A_{2}'$ & 1532.0&-70.3  \\
0,0,1,3,0,$A_{2}'$ & 1563.8&-38.5 \\
0,0,1,4,0,$A_{2}'$ & 1583.7&-18.7 \\
0,0,1,5,0,$A_{2}'$ & 1595.1&-7.2 \\
0,0,1,6,0,$A_{2}'$ & 1600.6&-1.7 \\
0,0,1,7,0,$A_{2}'$ & 1602.3&-0.01 \\
$D_{01}$\footnote{$^)$ Asymptotic energy of D$^-$+D$_{2}(v_{d}=0,j=1)$ dissociation}$^)$& 1602.3&\\
\hline
\end{tabular}
\caption{Energies (in cm$^{-1}$) of D$_3^-$ bound levels obtained in the present study. Energies in the second column are given with respect to $D_{as}$ (see Table \ref{tab:SummaryTab}, $D_{as}$  being the same for H$_3^-$ and D$_3^-$ ). The binding energies in the third column are given with respect to the dissociation limits $D_{00}$ for the $A_1'$ levels and $D_{01}$ for the $A_2'$  levels. The $D_{00}$ and $D_{01}$ dissociation D$_2$+D$^-$ limits are calculated with respect to $D_{as}$ and also given in the table.} 
\label{tab:D3_BoundStates}
\end{table}

\section{Permanent dipole moment surface for the H$_3^-$ ground state.}
\label{sec:dipole}

\begin{figure}[ht]
\includegraphics[width=12cm]{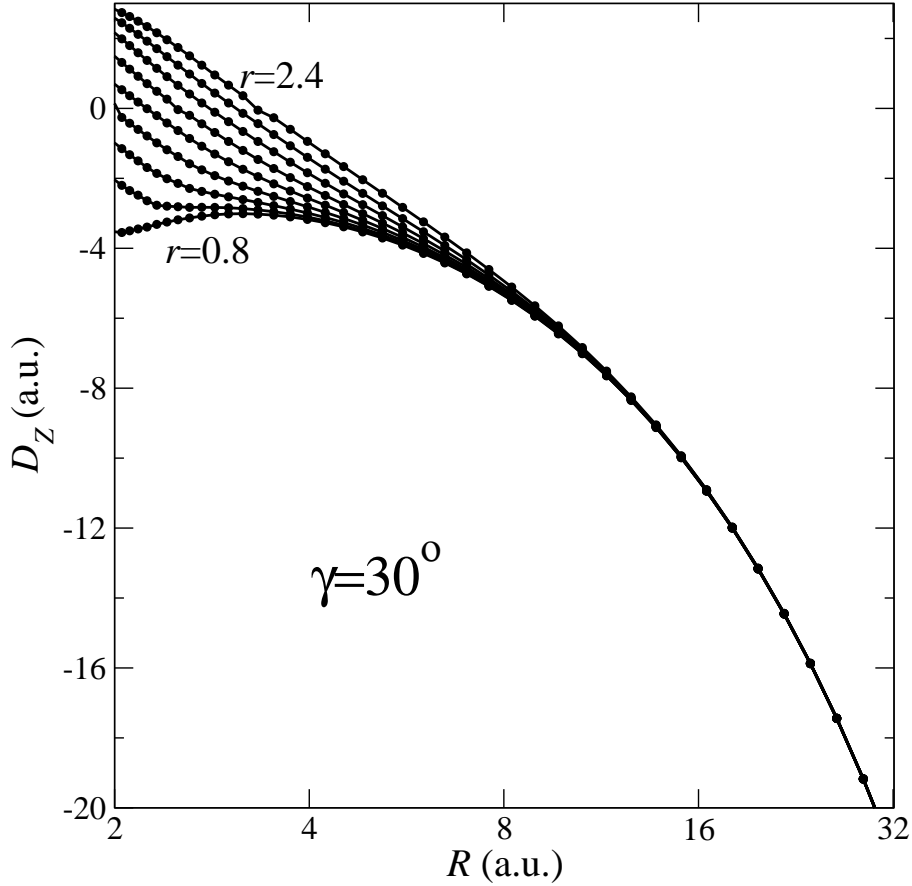}
\caption{The figure shows the $D_Z$ component of the permanent electric dipole of H$_3^-$ for one value of $\gamma=30$\textdegree~ and several values of $r$ from $r=$0.8 (the largest value of $D_Z$ at small $R$) to 2.4 (the smaller value of $D_Z$ at small $R$). The {\it ab initio} values and grid points are indicated by circles.}
\label{fig:Dz_gamma30}
\end{figure}
\begin{figure}[ht]

\includegraphics[width=12cm]{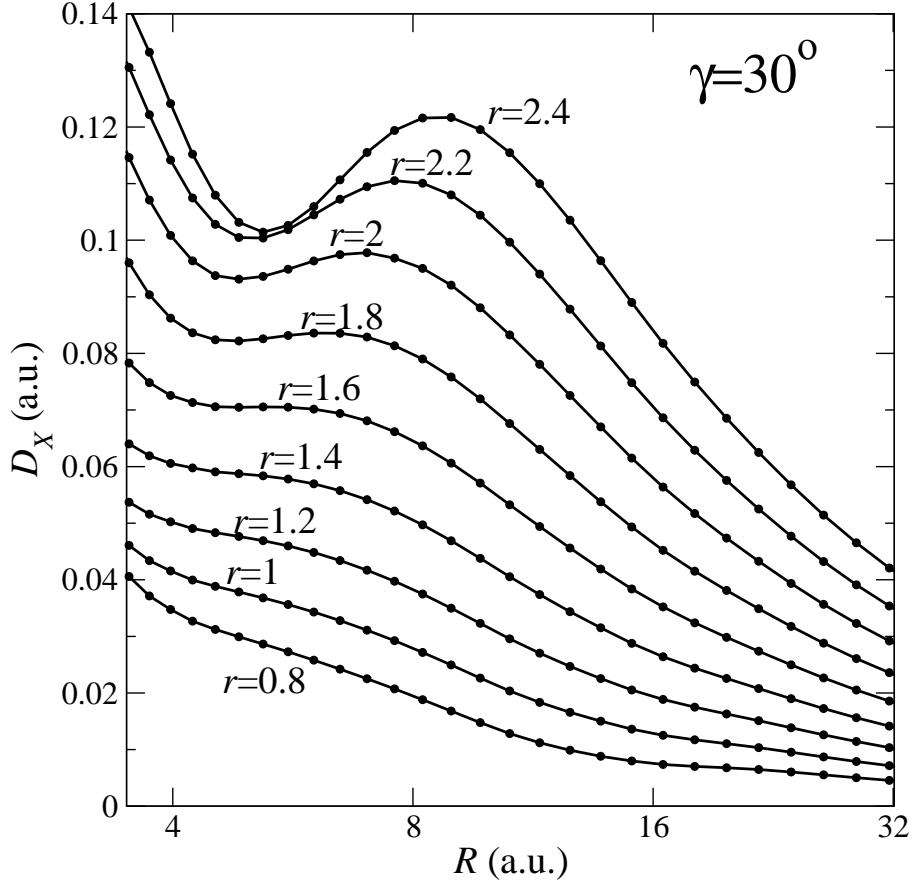}
\caption{The $D_X$ component of the permanent dipole of H$_3^-$ for  $\gamma=30$\textdegree~ and different  $r$. The {\it ab initio} values and grid points are indicated by circles. }
\label{fig:Dx_gamma30}
\end{figure}

The Molpro package delivers  also  the components of the permanent dipole moment together with the PES. For a fixed geometry, the obtained components of the dipole moment correspond to the coordinate system with the $X,Y,Z$ axes along the principal axes of inertia of the molecule. In our notations, the $Y$-axis is orthogonal to the plane of H$_3^-$, therefore $D_Y=0$. For $\gamma=0$ and 90\textdegree~ there is only one non-zero component ($D_Z$ in our notations). $D_Z$ becomes also equal to zero at equilateral geometries. For other geometries ($\gamma\ne$0 or 90\textdegree) there are two non-zero components, $D_Z$ and $D_X$. We used the following convention to label the components: at large values of $R$, the component $D_Z$ is the largest of the two in magnitude and negative, $D_X$ being the smallest (in magnitude) and positive. The choice of negative $D_Z$ at large $R$ corresponds to the $Z$ axis oriented from the center of mass of the molecule towards the H$^-$ ion. For small values of $R$ (for fixed $\gamma$ and $r$) the two components $D_Z$ and $D_X$  become comparable in magnitude. For such geometries they are identified by ensuring a smooth variation of the components with decreasing $R$. Notice that because the principal axes of inertia for H$_3^-$ and D$_3^-$ are the same, the obtained PDMS are the same for the two species. 

Figures \ref{fig:Dz_gamma30} and \ref{fig:Dx_gamma30} show {\it ab initio} values of the $D_Z$ and $D_X$ PDM components for $\gamma=30$\textdegree\ for  all nine values of $r=0.8,1,\cdots ,2.4$ a.u. as a function of $R$. As one can see from the figures, the sampling grid of {\it ab initio} geometries is dense enough to perform an interpolation procedure in order to calculate  $D_Z$ and $D_X$ at any arbitrary geometry. Therefore, in the central region in Fig. \ref{fig:Grid_Extrapolation}, we interpolate the $D_Z$ and $D_X$ PDMS using the same 3D B-spline procedure as for the PES interpolation. 

In region II, when $R\to\infty$, the analytical behavior of the largest component $D_Z$ is known: It decreases with increasing $R$ as (in a.u.) $D_Z \equiv er_\textrm{cm-H}=-2R/3$, where $r_\textrm{cm-H}$ is the distance between the center of mass of H$_3^-$ and the H$^-$ ion, and $e \equiv 1$ is the electron charge. The {\it ab initio} $D_Z$ values confirm this behavior, so the same formula is used for extrapolation of $D_Z$ in region II. Outside the central region in Fig. \ref{fig:Grid_Extrapolation}, the PDMS (except the $D_Z$ components in region II) are extrapolated using an empirical analytical formula. The empirical formula is obtained by inspecting the PDMS variations close to the boundaries of the {\it ab initio} region.  The smallest component $D_X$ varies at large $R$ as $k_x(r,\gamma)/R$, where $k_x(r,\gamma)$ is fixed by the value of the PES at the final point $R_f=56.4227$ a.u. of the grid, {\it i.e.}  $k_x=V(R_f;r,\gamma)R_f$. As previously, we obtain a surface for the $k_x(r,\gamma)$ function, which is computed at any geometry using a 2D B-spline interpolation. For the extrapolation of the PDMS in region I we found (empirically) a quadratic (for $D_Z$) and linear (for $D_X$) dependencies along $r$ with coefficients depending on $R$ and $\gamma$, which are evaluated from the corresponding boundary values of $D_Z$ and  $D_X$ respectively. Similarly, in region III and IV, we assumed a linear dependency along $r$ and $R$ respectively, with coefficients fixed by the boundary values of the PDMS.

\section{Summary and conclusions}
\label{sec:concl}
In the present study we have obtained the accurate potential energy surface and components of the permanent dipole moment for the H$_3^-$ van der Waals molecule. The surfaces were calculated on a dense grid of geometries that covers short, intermediate, and long-range regions. In total, the {\it ab initio} calculations were made for 3024 geometries. The large basis and grid used in the calculations allows us to suggest that the obtained PES is more accurate than the results of the previous study \cite{starck93,panda04}. Comparison of the long-range behavior of the obtained PES with the expected analytical behavior of the PES confirms this conclusion. No previous data on the dipole moment of H$_3^-$ is available.

The obtained  {\it ab initio} values for potential energy surface and the dipole moment components were used to construct  Fortran interpolation/extrapolation subroutines that calculate the energies and dipole moments for any arbitrary geometry. The subroutines interpolate the surfaces using B splines inside the box of {\it ab initio} geometries and extrapolate the surfaces outside of the box using analytical formulas based on the theoretical asymptotic behavior. The subroutines are available from the journal's EPAPS service. The energy surface can be used for all isotopologues of   H$_3^-$, the dipole moment surfaces in the present form can only be used for H$_3^-$ and D$_3^-$  isotopologues. For the H$_2$D$^-$ and D$_2$H$^-$ molecules, the dipole moments should be transformed to account for the different orientation of axes of inertia. Using the new potential surfaces, we have calculated the bound states for the H$_3^-$ and D$_3^-$ isotopologues.

A relatively large magnitude $|D|\sim 4$ a.u. of the dipole moment near equilibrium positions for bound vibrational states and a large size of the electronic clouds of H$^-$ suggest that the cross-section for the formation of H$_3^-$ stable molecules by the radiative association between H$_2$ and H$^-$  is significant. A relatively large dipole moment and existence of several bound levels suggests also that the  H$_3^-$ can be detected using the IR photoabsorption spectroscopy. 

The calculated energies of H$_3^-$ bound states can also be used to search for H$_3^-$ in the cold interstellar medium with a large fraction of ionized hydrogen (to have enough free electrons). We have developed a model for the formation of H$_3^-$ in radiative association collisions between H$_2$ and H$^-$. In the model, H$^-$ is formed by dissociative attachment of the electron to H$_2$. If a photoabsorption signal from H$_3^-$ (in mm range)  is detected, this would also be a signal for the presence of H$^-$ in the ISM:   H$^-$ itself cannot be detected directly. The details of the model as well as the calculated rates of radiative association collisions between H$_2$ and H$^-$ will be discussed in a forthcoming publication.

\begin{acknowledgments}
 The study was supported by the  {\it R\'eseau th\'ematique de recherches avanc\'ees "Triangle de la Physique"}, the National Science Foundation under grant PHY-0855622, and by an allocation of NERSC supercomputing resources. R.G. acknowledges  the generous support from {\it Insitut Francilien de recherches sur les atomes froids (IFRAF)}
\end{acknowledgments}

\end{document}